\documentclass[%
 reprint,
 groupedaddress,
 showpacs,preprintnumbers,
 nofootinbib,
 amsmath,amssymb,
 aps,
 pre,
 showkeys,
]{revtex4-1}
\usepackage{dcolumn}
\usepackage{latexsym}
\usepackage{amsmath}
\usepackage{amssymb}
\usepackage{amsthm}
\usepackage{dsfont}
\usepackage{longtable}
\usepackage{color}
\usepackage{verbatim}
\usepackage{graphicx}
\usepackage{epstopdf}
\usepackage{ulem}
\usepackage{cancel}

\usepackage[pdfstartview={FitH top},pdfpagemode=None,breaklinks]{hyperref}
\hypersetup{
    colorlinks,
    citecolor=black,
    filecolor=black,
    linkcolor=black,
    urlcolor=blue
}

\newcommand {\dd}[2] {\frac {\partial {#1} }{\partial {#2}}}

\newcommand{\Tr}{\mathop{\mathrm{Tr}}\nolimits}

\newcommand{\tg}{\mathop{\mathrm{tg}}\nolimits}
\newcommand{\arctg}{\mathop{\mathrm{arctg}}\nolimits}
\newcommand{\ch}{\mathop{\mathrm{ch}}\nolimits}
\newcommand{\ctg}{\mathop{\mathrm{ctg}}\nolimits}

\theoremstyle{definition}

\makeatletter
\@addtoreset{remark}{utv}
\@addtoreset{remark}{theorem}
\@addtoreset{paragraph}{section}
\makeatother

\allowdisplaybreaks
\begin{document}
\title[]
{Virial coefficients in $(\tilde{\mu},q)$-Bose gas model related to
 compositeness of particles and their interaction: temperature-dependence problem}

\author{A.M. Gavrilik}
\affiliation{Bogolyubov Institute for Theoretical Physics, Nat. Acad. of Sci. of Ukraine}
\address{14b, Metrolohichna Str., Kyiv 03680, Ukraine}
\email{omgavr@bitp.kiev.ua}
\author{Yu.A. Mishchenko}
\affiliation{Bogolyubov Institute for Theoretical Physics, Nat. Acad. of Sci. of Ukraine}
\address{14b, Metrolohichna Str., Kyiv 03680, Ukraine}


\begin{abstract}
We establish the relation of the second virial coefficient of
certain $(\tilde{\mu},q)$-deformed Bose gas model, recently proposed
by the authors in [Ukr. J. Phys., 2013], to the interaction and
compositeness parameters when  either of these factors is taken into
account separately. When the interaction is dealt with, the
deformation parameter becomes linked directly to the scattering
length, and the effective radius of interaction (in general, to
scattering phases). The additionally arising temperature dependence
is a new feature absent in the deformed Bose gas model within
adopted interpretation of the deformation parameters $\tilde{\mu}$
and $q$. Here the problem of the temperature dependence is analyzed
in detail and its possible solution is proposed.
\end{abstract}
\maketitle

\section{Introduction}

Nonlinear physical system involving either of the
nonideality/nonlinearity factors are often effectively described by
means of some deformed (algebraic, phenomenological or other) models
as counterpart to their corresponding ``ideal'' prototype. Deformed
Bose  gas models (DBGMs) along with deformed oscillators, deformed
quantum mechanics and some other extensions, being evolved since the
end of 1980s till now, belong to such models of an effective
description. Generally speaking the DBGMs may or may not be directly
linked with deformed oscillators, which in many cases are taken as
the structural object underlying the former. Soon after introducing
deformed oscillators or deformed
bosons~\cite{ArikCoon,Biedenharn_1989,Macfarlane_1989}, their use
for elaborating respective deformed analogs of Bose gas model became
popular. Already the early instances of DBGM, to list a
few~\cite{SuGe1993,MartinDelgado1991,*NeskovicUrosevic1992,Man'ko1993,AltherrGrandou1993,Chaichian1993,Ubriaco1997QGr},
had witnessed the appearance of very important new direction with a
long-term perspective and with good potential for useful (and
realistic) applications. The latter range from ${}^4He$
system~\cite{R-Monteiro} to e.g. such high-energy physics objects as
two- and three-particle correlations of pions generated in
relativistic collisions of
nuclei~\cite{Anchishkin,*Anchishkin2,*Anchishkin_Transverse,*Gavrilik_Sigma}.
These applications yielding a good effective description stimulated
the study of deeper reasons of the applicability of DBGMs to real
physical systems. Helpful, from this viewpoint, appears the idea
that deformation of ideal Bose gas model could and should provide an
efficient effective description of the properties of more realistic
(i.e. non-ideal) gases of Bose like particles. Moreover, the
deviation from strict ideality may originate for several reasons
(``nonideality'' factors). It was demonstrated~\cite{Avan} that
non-pointlike form of particles may serve as the first and most
obvious such a factor, and it is possible to link the parameter $q$
of deformation directly with the ratio of excluded volume (the sum
of nonzero proper volumes of the particles) to the whole volume.
  The next factor of non-ideality is the interaction between the
particles, and as shown in~\cite{Scarfone2009Int} it can be
naturally taken into account by a version of deformation.

More recently it became clear that the possibility of realization
(of operator algebra) of {\it composite} bosons by deformed bosons
proven in~\cite{GKM2} is naturally promoted to the elaboration of
DBGM able to effectively account for compositeness  of Bose
particles (the compositeness makes them quasibosons, differing from
strict bosons). At last, let us mention recent work~\cite{GM_Virial}
which shows how to incorporate simultaneously two different factors
of non-ideality: the compositeness of particles and their
interaction. That work motivated our present study.

Let us mention some versions of DBGMs and their  application to
physical systems in different contexts. Thermodynamics of the
$q$-DBGMs was studied e.g.
in~\cite{Lavagno_Thermostat,*Lavagno_Therm,Algin2010Fib,*Algin2012}; 
for the Bose condensation of the deformed gases
see~\cite{Salerno1994BEC,*Ubriaco1998BEC}. The DBGMs and many-body systems 
of $q$-bosons were applied to phonon gas in ${}^4He$ \cite{R-Monteiro}, to 
excitons in~\cite{Liu2001,*Zeng2011}, to a study of pairing 
correlations in nuclei~\cite{Sviratcheva}. Some of 
DBGMs were applied when studying two-particle correlation
functions~\cite{Man'ko1993Corr,Anchishkin,Adamska2004,Gavrilik_Sigma,GR_EPJA,GM_Exact}.
The extent or strength of deformation of the mentioned models
usually is characterized by one or more deformation parameters. Till
now, the question about the relation between the deformation
parameters and the microscopic nonideality factors and their
parameters remained opened, and the microscopic analysis of the
correspondence between a physical system and its deformed
counterpart was still absent.

In this work,  similarly to~\cite{GKM2} where the deformation
parameter for the realization of bifermionic composites
(quasibosons) by deformed bosons was related to the wavefunctions of
bifermionic states being realized, we establish the relation between
the deformation of a special (class of) DBGM and the characteristics
of the interaction along with compositeness of particles of a gas,
which the DBGM is implied to effectively incorporate jointly.
  As the criterion of the effective description (or realization) in the
former case~\cite{GKM2}, the realizability of quasibosonic operator
relations was taken. In the present case the proximity of the virial
expansion of the equation of state for non-ideal quantum gas to that
of the DBGM is chosen as such a criterion. The structure function
characterizing the DBGM of the effective description for the
concreteness is taken of the same special form as
in~\cite{GM_Virial}, however, the analogous consideration given
below may concern more general situation. This structure function is
the functional composition of those ones corresponding to effective
taking the compositeness and interaction into account each one
separately. The most optimal form of such a functional dependence is
also an open question.

Among central issues of this paper is the temperature dependence of
virial coefficients. According
to~\cite{SuGe1993,Gong1995,Ubriaco1998BEC,Lavagno_Therm,Scarfone2009Int,Algin2012,GR_Virial,*GR_UJP2013,GM_Virial}
the virial coefficients within the DBGMs studied therein depend only
on the deformation parameter(s) which in our interpretation are
interrelated with nonideality factors and thus should not depend on
the temperature. On the other hand, the virial coefficients for a
gas with interaction manifest temperature dependence~\cite{Pathria}.
Just this problem is in the focus of the present work.

\section{Relation of deformation parameters to the interaction between quasibosons and
to their compositeness}

We start with the recently obtained~\cite{GM_Virial} deformed virial expansion for the $\tilde{\mu},q$-deformed
Bose gas whose thermodynamics/statistical physics is given through the structure function
$\varphi_{\tilde{\mu},q}(N)$ (denote $[N]_q \equiv \frac{1-q^N}{1-q}$),
\begin{equation}\label{phi_mu_q}
\varphi_{\tilde{\mu},q}(N)\!=\!\varphi_{\tilde{\mu}}([N]_q) \!=\!
(1\!+\!\tilde{\mu}) [N]_q\!-\! \tilde{\mu} [N]^2_q.
\end{equation}
The structure function determines quantitatively how the
thermodynamics/statistical physics is ``deformed'' for that or
another system. Namely, in~\cite{GM_Virial} the $\tilde{\mu},q$-DBGM
 is constructed so that $\varphi_{\tilde{\mu},q}\bigl(z\frac{d}{dz}\bigr)$
replaces the derivative $z\frac{d}{dz}$ in the known relation for
the total number of particles given through partition function i.e.
$N = z\frac{d}{dz} \ln Z$, yielding the definition for the deformed
total number of particles in terms of the nondeformed partition
function:
\begin{equation}
\tilde{N} \equiv N^{(\tilde{\mu},q)}(z,V,T) \equiv \varphi_{\tilde{\mu},q}\Bigl(z\frac{d}{dz}\Bigr) \ln Z.
\end{equation}
All other deformed physical quantities are recovered using the (non-deformed) 
version of the relations of ideal quantum Bose gas. For instance, for the 
second virial coefficient, which is of interest for us herein, within the 
$\tilde{\mu},q$-DBGM we have obtained~\cite{GM_Virial}
\begin{equation}\label{V2(mu,q)}
V_2^{(\tilde{\mu},q)} = - \frac{\varphi_{\tilde{\mu},q}(2)}{2^{7/2}} = - \frac{(1+q)(1-\tilde{\mu}q)}{2^{7/2}}.
\end{equation}

In our treatment, the parameter $q$ of $\varphi_{\tilde{\mu},q}(N)$ corresponds 
to effective taking the interparticle interaction into account, and  $\tilde{\mu}$ 
-- to composite-structure effects. Somewhat earlier, the Arik-Coon structure 
function $[N]_q$ was used to effectively incorporate~\cite{Scarfone2009Int} 
the interaction between the particles of a gas of elementary bosons. 

Note that if, in addition to deformed thermodynamic relations, the structure function 
$\varphi_{\tilde{\mu},q}(N)$  describes some deformed boson algebra related to 
the $\tilde{\mu},q$-DBGM studied herein, certain ranges of admissible $\tilde{\mu}$ 
and $q$ hold. These can be deduced from the condition $\varphi_{\tilde{\mu},q}(n)\ge0$, 
$n=1..N_{\rm max}$ which corresponds to non-negativity of the norm of deformed 
boson Fock states ($N_{\rm max}$ is maximum occupation number). In particular the 
non-negativity of $\varphi_{\tilde{\mu},q}(2)$ yields $\tilde{\mu} q\le 1$ and $q\ge -1$. 
However we do not appeal to the relation with a deformed boson algebra.

Besides $\varphi_{\tilde{\mu},q}(N)$, one can take yet another versions of combining the two
structure functions $\varphi_{\tilde{\mu}}(N)$ and $[N]_q$, e.g. in the form $\varphi_{q,\tilde{\mu}}(N) \!=\![\varphi_{\tilde{\mu}}(N)]_q$,
or as the family with one more parameter: $t \varphi_{\tilde{\mu},q}(N)+ (1-t) \varphi_{q,\tilde{\mu}}$, for $0\le t\le1$.
Remark that the treatment below can be extended to the case of even more general structure
function $\varphi(N)$ when some of the deformation parameters are responsible for interparticle interaction,
and the others -- for the composite structure of particles in the effective description.

\paragraph{\bf Effective account for the particle-particle interaction
to $(\lambda^3/v)^2$-terms.}
As known, the deviation (from the ideal or non-interacting case) of
the second virial coefficient $V_2$ due to the two-particle
interaction is expressed  through the partial wave phaseshifts
$\delta_l(k)$ and the bound state (if any) energies $\varepsilon_B$
as follows~\cite{Pathria}
\begin{multline}\label{VirCoef}
V_2-V_2^{(0)} = -8^{1/2} \sum\nolimits_B e^{-\beta \varepsilon_B}-\\
- \frac{8^{1/2}}{\pi} \sum\nolimits_l' (2l+1) \int_0^\infty e^{-\beta\hbar^2k^2/m}
\frac{\partial\delta_l(k)}{\partial k} dk.
\end{multline}
Here $B$ runs over bound states, $l$ is the  angular momentum
quantum number and the summation is performed over even $l$ in
bosonic case, and over odd $l$ in the fermionic case. In low-energy
approximation we retain in~(\ref{VirCoef}) only the $l=0$ summand
($s$-wave approximation). The corresponding phaseshift $\delta_0(k)$ generally 
can be determined by solving Schrodinger equation for a specified interaction 
potential. However, in the low-energy limit (when $l=1$ effects are negligible) 
the following expansion known as effective range approximation 
holds~\cite{Newton1982Scattering,Flugge1999,Capri2002Nonrel}:
\begin{equation}\label{r_0}
k\, {\rm ctg}\, \delta_0\! =\! -\frac1a \!+\! \frac12 r_0 k^2 \!+\! ..., \ \ \ 
r_0 \!=\! 2\!\int\limits_0^\infty\! dr \Bigl[\Bigl(1\!-\!\frac{r}{a}\Bigr)^2\!\!-\!\chi_0^2(r)\Bigr],
\end{equation}
where $a$ is the scattering length, $r_0$ -- effective range (radius), and 
$\chi_0(r)$ being the radial wavefunction of the lowest state multiplied by $r$. 
Since for some typical potentials $r_0$ 
depends only on the range and depth of the potential, this expansion is 
sometimes called as ``shape-independent approximation''. For the shape-independent 
approximation we find $\dd{\delta_0}{k} = -a + (a-3r_0/2) a^2k^2 + O(k^4)$.
Putting this derivative in~(\ref{VirCoef}) and performing integration, 
within the $s$-wave approximation we obtain
\begin{multline}\label{VirCoef2}
V_2-V_2^{(0)} = -8^{1/2} \sum\nolimits_B e^{-\beta \varepsilon_B} + 2\frac{a}{\lambda_T}
- 2\pi^2 \Bigl(1-\frac32 \frac{r_0}{a}\Bigr) \Bigl(\frac{a}{\lambda_T}\Bigr)^3\\ + O((a/\lambda_T)^5),
\end{multline}
where $\lambda_T\equiv \lambda =h/\sqrt{2\pi m k_B T}$ is the thermal wavelength. Below we give
the explicit expressions for $V_2-V_2^{(0)}$, or for the pair $a$ and $r_0$ through which it
is expressed in~(\ref{VirCoef2}), for a number of potentials (their definitions and some details
are relegated to appendix~\ref{ap1}):
\newline $\bullet$ Hard spheres interaction potential~(\ref{HSpheres}). We have~\cite{Pathria}
\begin{equation}
V_2-V_2^{(0)} = 2\frac{D}{\lambda_T} + \frac{10\pi^2}{3} \Bigl(\frac{D}{\lambda_T}\Bigr)^5 + ... \ (l=0,2).
\end{equation}
\par\noindent$\bullet$ Constant repulsive potential~(\ref{CR-pot}). For this and subsequent potentials
the corresponding quantities are given using~\cite{Flugge1999}. So, we have
\begin{equation}\label{a,r_0-f}
a = R \Bigl(1-\frac{{\rm th}\, K_0R}{K_0R}\Bigr),\quad r_0=0.
\end{equation}
\par\noindent$\bullet$ Square-well potential~(\ref{SqW}). For this,
\begin{equation}
a = -R \Bigl(\frac{\tg K_0R}{K_0R}-1\Bigr),\ \ r_0 = R \Bigl(1-\frac{1}{K_0^2Ra}-\frac{R^2}{3a^2}\Bigr)
\end{equation}
\par\noindent$\bullet$ Anomalous scattering potential~(\ref{AnomScat}). For this,
\begin{equation}
a = R - \frac{{\rm th\,} K_0(R-r_1) + {\scriptstyle\frac{K_0}{K_1}} \tg K_1r_1}{K_0 \bigl(1 + {\scriptstyle\frac{K_0}{K_1}}
\tg (K_1r_1) {\rm th\,} K_0(R-r_1)\bigr)}.
\end{equation}
Somewhat awkward expression for $r_0$ is omitted.
\par\noindent$\bullet$ Scattering resonances~(\ref{SR_pot}). For this,
\begin{equation}
a = \frac{\Omega}{\Omega+1} R,\quad r_0 = \frac23 \frac{\Omega-1}{\Omega} R.
\end{equation}
\par\noindent$\bullet$ Modified P$\rm\ddot{o}$schl-Teller potential\,(\ref{PT-pot}). At integer $\lambda$,
\begin{equation}
a \!=\! \frac{1}{\alpha}\! \sum_{n=1}^{\lambda-1} \frac1n,\ \ r_0 \!=\! \frac{2}{3\alpha} \frac{\bigl(\sum\nolimits_{n=1}^{\lambda-1} n^{-1}\bigr)^3
- \sum\nolimits_{n=1}^{\lambda-1} n^{-3}}{\bigl(\sum\nolimits_{n=1}^{\lambda-1} n^{-1}\bigr)^2}.
\end{equation}
\par\noindent$\bullet$ Inverse power repulsive potential~(\ref{NegPow-pot}). For this,
\begin{equation}\label{a,r_0-l}
a = r_0 \frac{\Gamma(1-\frac{1}{2\eta})}{\Gamma(1+\frac{1}{2\eta})} \Bigl(\frac{g}{2\eta}\Bigr)^{\frac{1}{\eta}}, \ \ \
\eta= \frac{n-2}{2},
\end{equation}
and $r_0$ can be found from~(\ref{r_0}) using~(\ref{NP-chi0}).

With the data given above, we have the deviation of the second virial coefficient
in eq. (\ref{VirCoef2}) from that of ideal Bose gas, for each of
the considered potentials~(\ref{HSpheres}), (\ref{CR-pot}),...,(\ref{NegPow-pot}).

On the other hand, within $\varphi_{\tilde{\mu},q}$-deformed Bose gas model
we have~\cite{GM_Virial} (see also eq.~(\ref{V2(mu,q)})):
\begin{equation}\label{V(q)}
V^{(\tilde{\mu},q)}_2\!-\!V_2^{(0)}|_{\tilde{\mu}=0} \!\!=\!\! \frac{2-\varphi_{\tilde{\mu},q}(2)}{2^{7/2}}|_{\tilde{\mu}=0} \!\!=\!\! \frac{1-q}{2^{7/2}}.
\end{equation}
By juxtaposing this with~(\ref{VirCoef2}), we obtain
\begin{multline}\label{q(a,r,T)}
q\!=\!q(a,r_0,T) \!=\! 1-2^{9/2} \frac{a}{\lambda_T} + 2^{9/2} \pi^2 \Bigl(1-\frac32 \frac{r_0}{a}\Bigr) \Bigl(\frac{a}{\lambda_T}\Bigr)^3+...\\
+2^5 \!\sum\nolimits_B \!e^{-\beta \varepsilon_B},
\end{multline}
which constitutes one of our main results. Of course, this formula should be appended
with $a$ and $r_0$ taken e.g. for the chosen cases from (\ref{a,r_0-f})-(\ref{a,r_0-l}),
or for any other desired case.

The temperature dependence of the deformation parameter in (\ref{q(a,r,T)}) appears somewhat
unexpected since, in our interpretation, the deformation parameter characterizes
the nonideality of deformed Bose gas model as a whole, and $T$ is its internal parameter.
One of the approaches to resolve this issue consists
in a modification of the very deformation in deformed Bose gas model. For instance, we can use
the extended deformed derivative (here $z=e^{\beta\mu}$ is the fugacity, $\mu$ the chemical potential)
\begin{equation}
z\dd{}{z} \rightarrow z\tilde{\mathcal{D}}_z\equiv \varphi\Bigl(z\dd{}{z}\Bigr) + \chi\Bigl(z\dd{}{z}\Bigr) \dd{}{\beta} + g(\beta) \rho\Bigl(z\dd{}{z}\Bigr), \label{modif}
\end{equation}
with structure functions $\varphi$, $\chi$, $\rho$, in the relation
\begin{multline*}
\tilde{N} \!=\! z \tilde{\mathcal{D}}_z \ln Z^{(0)} \!=\\
= \varphi({\scriptstyle z\dd{}{z}}) \ln Z^{(0)} - \chi({\scriptstyle z\dd{}{z}}) U^{(0)} 
- \beta g(\beta) \rho({\scriptstyle z\dd{}{z}}) \Phi_G^{(0)}.
\end{multline*}
Here $Z^{(0)}$, $U^{(0)}$ and $\Phi_G^{(0)}$ are nondeformed partition function, 
internal energy and Gibbs thermodynamic potential respectively; $(z,V,T)$ serve as 
independent variables. Thus, on the thermodynamics level, $\chi$ and $\rho$ 
reflect the effect on the total number of particles of the internal energy and Gibbs 
thermodynamic potential which now appear on the same footing as the logarithm of 
grand partition function. The corresponding analysis will be carried out in
sec.~\ref{sec:modif_der} below. 

\noindent{\it Remark.} It is worth to estimate the relative magnitude of the terms $-8^{1/2} \sum\nolimits_B e^{-\beta \varepsilon_B}$ and
$2\frac{a}{\lambda_T}$ in~(\ref{VirCoef2}) at low-energy scattering when bound states do exist.
According to~\cite{Flugge1999} we have the following estimate for the binding energy in terms of scattering data:
\begin{equation*}
\varepsilon_B \simeq - \frac{\hbar^2}{2ma^2} (1+\frac{r_0}{a}).
\end{equation*}
Using this we come to
\begin{multline}
-8^{1/2} e^{-\beta \varepsilon_B} + 2\frac{a}{\lambda_T} \simeq -8^{1/2} e^{\frac{\hbar^2}{2ma^2 k_B T}\bigl(1+\frac{r_0}{a}\bigr)} + 2\frac{a}{\lambda_T} = \\
\shoveleft{= -8^{1/2} e^{\frac{1}{4\pi} \frac{\lambda_T^2}{a^2} (1+r_0/a)} + 2\frac{a}{\lambda_T} < 0
\ \ \ \text{for}\ \ a/\lambda_T<1.}
\end{multline}
Thus, the binding energy term in~(\ref{VirCoef2}) (if a bound state exists) is dominating over
$2\frac{a}{\lambda_T}$ for small $a/\lambda_T$.

\paragraph{\bf Effective account for the compositeness of particles up to $(\lambda^3/v)^2$-terms.}

Let us now evaluate the second virial coefficient in the absence of
explicit interaction between quasibosons (composite bosons).
 Note that the partition function from which the second virial
coefficient can be extracted, for the system of composite bosons
 within a general framework was considered in~\cite{Shiau2013Partition}.
 Within our approach (which is both effective and efficient), however,
 the task of obtaining the virial coefficient(s) is completely tractable
 leading for the deformed Bose gas to exact results.

Two-component quasibosons concerned here have the following
creation/annihilation operators~\cite{Avan,GKM2}
\begin{equation} \label{anzats}
A^{\dag}_{\alpha}=\sum\limits_{\mu\nu}\Phi^{\mu\nu}_{\alpha}a^{\dag}_{\mu}b^{\dag}_{\nu},\quad
A_{\alpha}=\sum\limits_{\mu\nu}\overline{\Phi}^{\mu\nu}_{\alpha}b_{\nu}a_{\mu},
\end{equation}
where $a^\dag_\mu$, $b^\dag_\nu,$ and $a_\mu$, $b_\nu$ are the creation and annihilation operators for
the constituent fermions, and the set of matrices $\Phi_\alpha^{\mu\nu}$ determine the quasiboson wavefunction.
As a starting point we take the known general expression for 2nd virial coefficient~\cite{Pathria}
\begin{equation}\label{V2_def}
V_2=\frac{1}{2!V}[({\rm Tr}_1\, e^{-\beta H_1})^2-{\rm Tr}_2\, e^{-\beta H_2}]. 
\end{equation}
   Here ${\rm Tr}_1$ denotes the trace over one-quasiboson states and
${\rm Tr}_2$ -- over the states of two quasibosons; $H_1$ and $H_2$
are respectively one- and two-quasibosonic Hamiltonians. The
distinction between the second virial coefficients for the ideal
Bose- and ideal Fermi gases is caused by the nilpotency of the
fermionic creation operators, and  consequently by the nullifying of
the respective terms in ${\rm Tr}_2\, e^{-\beta H_2}$
from~(\ref{V2_def}). Analogously, in the case of bi-fermionic
quasibosons the nonzero summands from ${\rm Tr}_2\, e^{-\beta H_2}$
are determined by the condition
$|(A_\alpha^\dag)^2|0\rangle|^2\ne0$. Let us calculate
$|(A_\alpha^\dag)^2|0\rangle|^2$:
\begin{multline}
|(A_\alpha^\dag)^2|0\rangle|^2 =\!\!\!\!\!\sum_{\mu_1\mu_2..\nu_1'\nu_2'}\!\!\!\!\! \langle 0|b_{\nu_2'}a_{\mu_2'}b_{\nu_1'}a_{\mu_1'}
\overline{\Phi_\alpha^{\mu_1'\nu_1'}}\, \overline{\Phi_\alpha^{\mu_2'\nu_2'}} \Phi_\alpha^{\mu_1\nu_1} \Phi_\alpha^{\mu_2\nu_2}\cdot\\
\cdot a^\dag_{\mu_1} b^\dag_{\nu_1} a^\dag_{\mu_2} b^\dag_{\nu_2} |0\rangle = 2 \!\!\!\!\!\sum_{\mu_1\!\ne\! \mu_2, \nu_1\!\ne\! \nu_2}\!\!\!\!\! \bigl(|\Phi_\alpha^{\mu_1\nu_1}|^2 |\Phi_\alpha^{\mu_2\nu_2}|^2
- \overline{\Phi_\alpha^{\mu_1\nu_2}}\, \overline{\Phi_\alpha^{\mu_2\nu_1}}\cdot\\
\cdot \Phi_\alpha^{\mu_1\nu_1} \Phi_\alpha^{\mu_2\nu_2}\bigr)
= 2\bigl(1-\Tr (\Phi_\alpha \Phi_\alpha^\dag \Phi_\alpha \Phi_\alpha^\dag)\bigr).
\end{multline}
The traces in~(\ref{V2_def}) are calculated as follows:
\begin{equation}\label{Tr1}
{\rm Tr}_1\, e^{-\beta H_1} = \sum_{{\bf k}_1 n_1} \langle0|A_{{\bf k}_1 n_1} e^{-\beta H_1}A^\dag_{{\bf k}_1 n_1}|0\rangle
= \sum_{{\bf k}_1 n_1} e^{-\beta\varepsilon_{{\bf k}_1 n_1}},
\end{equation}
\begin{multline}\label{Tr2}
{\rm Tr}_2\, e^{-\beta H_2} \!=\! 1/2 \!\!\!\!\!\sum_{({\bf k}_1 n_1)\ne ({\bf k}_2 n_2)}\!\!\!\!\! \langle0| A_{{\bf k}_2 n_2} A_{{\bf k}_1 n_1} e^{-\beta H_2}\! A^\dag_{{\bf k}_1 n_1}\! A^\dag_{{\bf k}_2 n_2}|0\rangle\\
+ \frac{1}{|(A_{{\bf k}_1 n_1}^\dag)^2|0\rangle|^2} \mathop{\sum\nolimits^{\prime}}\limits_{{\bf k}_1 n_1} \langle0|(A_{{\bf k}_1 n_1})^2 e^{-\beta H_2} (A^\dag_{{\bf k}_1 n_1})^2 |0\rangle =\\
= \frac12 \Bigl(\sum_{{\bf k}_1 n_1} e^{-\beta\varepsilon_{{\bf k}_1 n_1}}\Bigr)^2 - \frac12 \sum_{{\bf k}_1 n_1} e^{-2\beta\varepsilon_{{\bf k}_1 n_1}} + \mathop{\sum\nolimits^{\prime}}\limits_{{\bf k}_1 n_1} e^{-2\beta\varepsilon_{{\bf k}_1 n_1}}
\end{multline}
Here ${\bf k}_{1,2}$ is the momentum quantum number, $n_{1,2}$ contains all the other quasibosonic
quantum numbers, $\varepsilon_{{\bf k}_1 n_1}$ is the energy of quasiboson in the state $|{\bf k}_1 n_1\rangle$
and the prime in $\sum'$ implies the summation over all the modes $({\bf k},n)$ for which $(A_{{\bf k},n}^\dag)^2|0\rangle\ne0$.
Substituting~(\ref{Tr1}) and (\ref{Tr2}) in (\ref{V2_def}) and splitting $\varepsilon_{{\bf k} n}$ into
kinetic energy $\frac{\hbar^2{\bf k}^2}{2m}$ and internal energy $\varepsilon_n^{int}$ as $\varepsilon_{{\bf k} n}
= \frac{\hbar^2{\bf k}^2}{2m} + \varepsilon_n^{int}$ we obtain
\begin{equation}
V_2(T) = \frac{1}{2^{5/2}} \sum_n e^{-2\beta \varepsilon_n^{int}} - \frac{\lambda_T^3}{V} \mathop{\sum\nolimits^{\prime}}\limits_{{\bf k} n}
e^{-2\beta \bigl(\frac{\hbar^2{\bf k}^2}{2m} + \varepsilon_n^{int}\bigr)}.
\end{equation}
If for all the $({\bf k},n)$-modes $(A_{{\bf k},n}^\dag)^2|0\rangle\ne0$, then performing the
summation over $\bf k$ according to $\sum_{\bf k} e^{-2\beta \frac{\hbar^2{\bf k}^2}{2m}} = 2^{-3/2} V/\lambda_T^3$ we obtain
\begin{equation}\label{V2-V^0}
V_2(T) - V_2^{(0)} = - \frac{1}{2^{5/2}} \Bigl(\sum_n e^{-2\beta \varepsilon_n^{int}} - 1\Bigr).
\end{equation}

On the other hand, in the deformed case we have
 the (exact) result~\cite{GM_Virial} (see also eq.
 (\ref{V2(mu,q)})), i.e.
\begin{equation}\label{V(mu)}
V^{(\tilde{\mu},q)}_2\!-\!V_2^{(0)}|_{q=1} \!=\!
\frac{2-\varphi_{\tilde{\mu},q}(2)}{2^{7/2}}|_{q=1} \!=\! \frac{\tilde{\mu}}{2^{5/2}}
\end{equation}
from which after juxtaposing, according to our interpretation, with (\ref{V2-V^0}) we arrive at
\begin{equation}\label{mu(E,T)}
\tilde{\mu}=\tilde{\mu}(\varepsilon_n^{int}, \Phi^{\mu\nu}_{\alpha}, T) = 1 - \sum_n e^{-2\beta \varepsilon_n^{int}}
\end{equation}
(the dependence on $\Phi^{\mu\nu}_{\alpha}$ is retained for general case). As now is seen, the
obtained difference~(\ref{V2-V^0}) is mainly related with the internal
energy of a quasiboson, not with its (nonbosonic) commutation relations.

The structure function $\varphi_{\tilde{\mu},q}(N)$ with
$q\!=\!q(a,r_0,T)$,
$\tilde{\mu}\!=\!\tilde{\mu}(\varepsilon_n^{int},
\Phi^{\mu\nu}_{\alpha}, T)$ is chosen for the goal of the effective
account (in certain approximation) for the factors of interaction
and of composite structure of particles of a gas.
  Let us emphasize that the direct microscopic treatment may lead to
  quite different relation between the second virial coefficient
  incorporating the both factors (interaction and compositeness) and
  the virial coefficients involving only one nontrivial factor.
  The functional composition as in~(\ref{phi_mu_q}) may not already hold,
nevertheless, the linear part of the Taylor expansion of
$V_2^{(\tilde{\mu},q)}$ in small $\epsilon = q-1$ and $\tilde{\mu}$ may
coincide with the corresponding part found from the microscopic
treatment.

It is clear that the modification of deformation according to~(\ref{modif}) may
lead to quite different dependence of deformation parameters on the characteristics
of interaction and compositeness.

Let us note that the major deformation structure function $\varphi$
in~(\ref{modif}) is a general one. The choice $\varphi\Bigl(z\dd{}{z}\Bigr) =
\varphi_{\tilde{\mu},q}\Bigl(z\dd{}{z}\Bigr)$ results in the formulas (\ref{V(q)}) and 
(\ref{V(mu)}) for the virial coefficient $V_2$. Clearly, other choices for $\varphi$ 
in (\ref{modif}) will result in other form of respective virial coefficient $V_2$ and 
the respective temperature dependence.

\section{Modification of derivative $z\frac{d}{dz}$ aimed to yield temperature
dependent virial coefficients}\label{sec:modif_der}

As already mentioned, we can obtain temperature dependent deformed
(i.e. within the deformation-based approach) virial coefficients say
by performing the extension of the deformed derivative,
see~(\ref{modif}). The functions $\varphi$, $\chi$, $\rho$ should
not depend on the temperature. The term $g(\beta)
\rho\Bigl(z\dd{}{z}\Bigr)$ is introduced in order to reflect the
ambiguity in the (left or right) position of
$\partial/\partial\beta$, i.e. to cover the terms like $\dd{}{\beta}
\chi\Bigl(z\dd{}{z}\Bigr)$. This can be verified by means of the
commutation relation
\begin{equation}
[\partial/\partial\beta,f(z\partial/\partial z)]= - \beta^{-1} \, (z\partial/\partial z)\cdot f'(z\partial/\partial z).
\end{equation}
The noncommutativity of derivatives $\partial/\partial\beta$ and $z\partial/\partial z$ is observed
after presenting $z\partial/\partial z$ as $\beta^{-1} \partial/\partial \mu$, where $\mu$ is chemical potential
(recall that $z=e^{\beta\mu}$).

Applying deformed derivative~(\ref{modif}) to the known expansion for the partition function
\begin{equation*}
\ln Z^{(0)}(z,V,T) = \frac{V}{\lambda_T^3} \sum_{n=1}^\infty \frac{z^n}{n^{5/2}}
\end{equation*}
we obtain the following series for the deformed
(that is why we use tilde) total number of particles
\begin{multline}\label{N_deform}
\tilde{N} \!=\! z\tilde{\mathcal{D}}_z \ln Z^{(0)}(z,V,T) \!=\! \frac{V}{\lambda_T^3} \sum_{n=1}^\infty \bigl[\varphi(n) \!+\!\beta^{-1}\bigl(n\chi(n) \ln z-\\
-\!3/2\chi(n)\!+\! n\chi'(n)\bigr) \!+\! g(\beta)\rho(n)\bigr] \frac{z^n}{n^{5/2}}.
\end{multline}
Deformed partition function is then recovered as
\begin{multline}\label{Z_deform}
\ln \tilde{Z} \!=\! \bigl(d/dz\bigr)^{-1} \tilde{N} \!=\! \frac{V}{\lambda_T^3} \sum_{n=1}^\infty \bigl[\varphi(n) \!+\! \beta^{-1}\bigl(n\chi(n)\ln z\!-\\
-\!5/2\chi(n)\!+\!n\chi'(n)\bigr) \!+\! g(\beta)\rho(n)\bigr] \frac{z^n}{n^{7/2}}.
\end{multline}
Expanding fugacity as $z=z_0 + z_1 \frac{\lambda_T^3}{\tilde{v}} + z_2 \bigl(\frac{\lambda_T^3}{\tilde{v}}\bigr)^2 + ...$,
denoting by $\tilde{v}=\frac{\tilde{N}}{V}$ the deformed specific volume, and remembering that $z_i=z_i(T)$, $i=0,1,...$,
after substituting the resulting expansion into~(\ref{N_deform}) we obtain the relation
\begin{equation}
\frac{\lambda_T^3}{\tilde{v}} = \sum_{n=0}^\infty R_n(T;\varphi,\chi,\rho) \Bigl(\frac{\lambda_T^3}{\tilde{v}}\Bigr)^n
\end{equation}
where the coefficients at the same powers of $\frac{\lambda_T^3}{\tilde{v}}$ in the l.h.s. and r.h.s.
should be
\begin{align}
&R_0\!\equiv\! \sum_{n=1}^\infty \bigl[\varphi(n) \!+\! \beta^{-1}\bigl(n\chi(n)\ln z_0\!-\!3/2 \chi(n)\!+\! n\chi'(n)\bigr) +\nonumber\\
&\qquad+\!g(\beta)\rho(n)\bigr] \frac{z_0^n}{n^{5/2}} \!=\! 0,\label{R0}\\
&R_1\!\equiv \!z_1\! \sum_{n=1}^\infty \bigl[\varphi(n) \!+\! \beta^{-1}\!\bigl(n\chi(n)\ln z_0\!-\!1/2\chi(n)\!+\! n\chi'(n)\bigr) +\nonumber\\
&\qquad+\!g(\beta)\rho(n)\bigr] \frac{z_0^{n-1}}{n^{3/2}} \!=\! 1,\\
&R_2\!\equiv\! \Bigl(\frac{z_2}{z_1}\!-\!\frac12 \frac{z_1}{z_0}\Bigr) P_1 + \frac12 \frac{z_1^2}{z_0} \sum_{n=1}^\infty \bigl[\varphi(n) \!+\! \beta^{-1}\bigl(n\chi(n)\ln z_0+\!\nonumber\\
&\qquad+1/2\chi(n)\!+\! n\chi'(n)\bigr) \!+\! g(\beta)\rho(n)\bigr] \frac{z_0^{n-1}}{n^{1/2}} \!=\! 0,\\
&\hspace{30mm}\,.\,.\,.\,.\,.\ .\nonumber
\end{align}
Similarly, for the deformed equation of state $\frac{\tilde{P}V}{k_B T} = \ln \tilde{Z}$, using~(\ref{Z_deform})
we find the following virial $\lambda_T^3/\tilde{v}$-expansion
\begin{equation}\label{VirExp}
\frac{\tilde{P}}{k_B T} \!=\! \frac{1}{\lambda_T^3} \tilde{V}_0(T;\varphi,\chi,\rho) + \tilde{v}^{-1} \sum_{n=1}^\infty
\tilde{V}_n(T;\varphi,\chi,\rho) \Bigl(\frac{\lambda_T^3}{\tilde{v}}\Bigr)^{n-1}
\end{equation}
with virial coefficients
\begin{align}
&\tilde{V}_0 \!\equiv \sum_{n=1}^\infty \bigl[\varphi(n) \!+\! \beta^{-1}\bigl(n\chi(n)\ln z_0\!-\!5/2\chi(n)\!+\!n\chi'(n)\bigr) +\nonumber\\
&\qquad+\!g(\beta)\rho(n)\bigr] \frac{z_0^n}{n^{7/2}} \!=\! 0,\label{V0}\\
&\tilde{V}_1\!\equiv \!z_1\! \sum_{n=1}^\infty \bigl[\varphi(n) \!+\! \beta^{-1}\!\bigl(n\chi(n)\ln z_0\!-\!3/2\chi(n)\!+\! n\chi'(n)\bigr) + \nonumber\\
&\qquad+\!g(\beta)\rho(n)\bigr] \frac{z_0^{n-1}}{n^{5/2}} \!=\! 1,\label{V1}\\
&\tilde{V}_2\!\equiv\! \Bigl(\frac{z_2}{z_1}\!-\!\frac12 \frac{z_1}{z_0}\Bigr) \tilde{V}_1 + \frac12 \frac{z_1^2}{z_0} \sum_{n=1}^\infty \bigl[\varphi(n) \!+\! \beta^{-1}\bigl(n\chi(n)\ln z_0-\nonumber\\
&\qquad- \!1/2\chi(n) \!+\! n\chi'(n)\bigr) \!+\! g(\beta)\rho(n)\bigr] \frac{z_0^{n-1}}{n^{3/2}},\\
&\hspace{30mm}\,.\,.\,.\,.\,.\ .\nonumber
\end{align}
The equalities in~(\ref{V0}), (\ref{V1}) are imposed in order that virial expansion~(\ref{VirExp}) reproduces
the corresponding limit of classical ideal gas. One of the solutions of~(\ref{R0}) and (\ref{V0})
is $z_0=0$. Let us dwell on this case. Deformed equation of state $\tilde{P}=\tilde{P}(\lambda_T^3/\tilde{v})$,
see~(\ref{VirExp}), can be written in the implicit parametric form (see~(\ref{N_deform}), (\ref{Z_deform})):
\begin{align}
&\frac{\lambda_T^3}{\tilde{v}} \!=\! \sum_{n=1}^\infty \bigl[\varphi(n) \!+\! \beta^{-1}\bigl(n\chi(n)\ln z\!-\!3/2\chi(n) \!+\!n\chi'(n)\bigr) +\nonumber\\
&\qquad+\! g(\beta)\rho(n)\bigr] \frac{z^n}{n^{5/2}}\label{lambda3/v},\\
&\frac{\tilde{P}}{k_B T} \!=\! \frac{1}{\lambda_T^3} \sum_{n=1}^\infty \bigl[\varphi(n) \!+\! \beta^{-1}\bigl(n\chi(n)\ln z\!-\!5/2\chi(n) \!+\!n\chi'(n)\bigr)\nonumber\\ &\qquad+\! g(\beta)\rho(n)\bigr] \frac{z^n}{n^{7/2}}\label{P/kT}.
\end{align}
Value $z=z_0=0$ corresponds to $\lambda_T^3/\tilde{v}|_{z=0}=0$, as $z \ln z\rightarrow 0$ at $z\to 0$ in~(\ref{lambda3/v}).
Consider the first derivative of~(\ref{P/kT}) by $\lambda_T^3/\tilde{v}$ namely
\begin{multline*}
\dd{\bigl(\tilde{P}/(k_B T)\bigr)}{\bigl(\lambda_T^3/\tilde{v}\bigr)} = \frac{\partial\bigl(\tilde{P}/(k_B T)\bigr)/\partial z}{\partial\bigl(\lambda_T^3/\tilde{v}\bigr)/\partial z} = \frac{1}{\lambda_T^3}\cdot\\
\cdot \frac{\sum\limits_{n=1}^\infty\! \bigl[\varphi(n) \!+\! \beta^{-1}\!\bigl((n \ln z\!-\!{\scriptstyle\frac32}) \chi(n)\!+\!n\chi'(n)\bigr) \!+\! g(\beta)\rho(n)\bigr] \frac{z^n}{n^{5/2}}}{\sum\limits_{n=1}^\infty \!\bigl[\varphi(n) \!+\! \beta^{-1}\!\bigl((n \ln z\!-\!{\scriptstyle\frac12})\chi(n)\!+\!n\chi'(n)\bigr) \!+\! g(\beta)\rho(n)\bigr] \frac{z^n}{n^{3/2}}}\\
\mathop{\rightarrow}\limits_{z\to 0}\! \frac{1}{\lambda_T^3} \frac{\varphi(1) \!+\! \beta^{-1}\! \bigl(\chi(1) (\ln z\!-\!{\scriptstyle\frac32})\!+\!\chi'(1)\bigr) \!+\! g(\beta)\rho(1)}{\varphi(1) \!+\! \beta^{-1}\! \bigl(\chi(1) (\ln z\!-\!{\scriptstyle\frac12})\!+\!\chi'(1)\bigr) \!+\! g(\beta)\rho(1)} \!\mathop{\rightarrow}\limits_{z\to 0}\! \frac{1}{\lambda_T^3}.
\end{multline*}
For the second derivative we obtain
\begin{multline}
\dd{^2\bigl(\tilde{P}/(k_B T)\bigr)}{\bigl(\lambda_T^3/\tilde{v}\bigr)^2} \!=\! \frac{\dd{^2\bigl(\tilde{P}/(k_B T)\bigr)}{z^2} \dd{(\lambda_T^3/\tilde{v})}{z} \!-\! \dd{\bigl(\tilde{P}/(k_B T)\bigr)}{z} \dd{^2(\lambda_T^3/\tilde{v})}{z^2}}{\bigl(\partial(\lambda_T^3/\tilde{v})/\partial z\bigr)^3}\\
\shoveleft{= \lambda_T^{-3} [\varphi(1) + \beta^{-1} (\chi(1) (\ln z\!-\!1/2)\!+\!\chi'(1)) + g(\beta)\rho(1)]^{-3} \cdot}\\
\shoveleft{\bigl[\beta^{-2} \chi^2(1) z^{-1} - 2^{-3/2} \beta^{-2} \chi(2)\chi(1) \ln^2 z + 2^{-5/2} \beta^{-1}\cdot}\\
\bigl(-\varphi(2)\chi(1) + 5/2 \beta^{-1} \chi(2)\chi(1) - g(\beta) \chi(1)\rho(2) -2\varphi(1)\chi(2) -\\
- 2 g(\beta)\chi(2)\rho(1) - 2\beta^{-1} \chi(1)\chi'(2) - 2\beta^{-1} \chi'(1)\chi(2)\bigr) \ln z -\\
\shoveleft{- 2^{-5/2} \bigl(\varphi(2)\varphi(1) - 5/2 \beta^{-1} \varphi(2)\chi(1) + g(\beta) \varphi(2)\rho(1) +}\\
+ 5/2 \beta^{-1} \varphi(1)\chi(2) + g(\beta)\varphi(1)\rho(2) - 17/4 \beta^{-2} \chi(1)\chi(2) +\\
+ 5/2 \beta^{-1}\!g(\beta) \chi(2)\rho(1) - 5/2 \beta^{-1}\!g(\beta) \chi(1)\rho(2) + g^2(\beta)\rho(1)\rho(2)\\
+ \beta^{-1} \varphi(2)\chi'(1) + 2\beta^{-1} \varphi(1) \chi'(2) + 2 \beta^{-2} \chi'(1)\chi'(2) + \\
+ 5/2 \beta^{-2} \chi(2)\chi'(1) - 4 \beta^{-2} \chi'(2)\chi(1) + 2 \beta^{-1} g(\beta) \chi'(2)\rho(1) +\\
+ \beta^{-1} g(\beta) \chi'(1)\rho(2)\bigr)\bigr].
\end{multline}
As it is seen from the last expression, for the second deformed virial coefficient
$\tilde{V}_2 = \frac12 \lambda_T^3  \dd{^2(\tilde{P}/(k_B T))}{(\lambda_T^3/\tilde{v})^2}$
to be finite at $z\to z_0=0$ we have to require $\chi(1)=0$. Then
\begin{equation*}
\tilde{V}_2 \!=\! -\! \frac{2\beta^{-1} \!\chi(2)\ln z \!+\! \varphi(2)\!+\!\beta^{-1}\! ({\scriptstyle\frac52} \chi(2)\!+\!2\chi'(2)) \!+\! g(\beta)\rho(2)}{2^{7/2}\, \bigl(\varphi(1) \!+\! \beta^{-1} \!\chi'(1) \!+\! g(\beta) \rho(1)\bigr)^2}.
\end{equation*}
Likewise, the requirement of finiteness leads to $\chi(2)=0$ and  thus to
\begin{equation}
\tilde{V}_2 = -\frac{1}{2^{7/2}} \frac{\varphi(2) + 2\beta^{-1} \chi'(2)+ g(\beta)\rho(2)}{(\varphi(1) + \beta^{-1}\chi'(1) + g(\beta)\rho(1))^2}.
\end{equation}
The obtained general formula involves dependence on the choice of
deformation (through the values $\varphi(k)$, $\chi'(k)$, $\rho(k)$, $k\!=\!\overline{1,2}$
of the structure functions $\varphi,\,\chi,\,\rho$ from (\ref{modif})).

In some situations it may be more convenient to deal with virial
$z$-expansions. Say, for the total number of particles we have
\begin{multline}\label{N(z)}
N(z,V,T) = \frac{V}{\lambda_T^3} \Bigl[z + 2 \Bigl(\frac{1}{2^{5/2}} - 2\frac{a}{\lambda_T}\Bigr)z^2 + ...\Bigr] \simeq\\
\simeq z\tilde{\mathcal{D}}_z \Bigl\{\frac{V}{\lambda_T^3} \Bigl(z + \frac{1}{2^{5/2}} z^2 + ...\Bigr)\Bigr\}.
\end{multline}
Using the last expression we can compare the result of the
microscopic treatment with the action of the deformation.
 In the r.h.s. of~(\ref{N(z)}) we have exactly the r.h.s.
of~(\ref{N_deform}). Taking there $\chi(n)=0$, $n=1,2,...$ (as the
simplest variant to exclude singularity at $z\to z_0=0$) and
comparing the first two terms with the corresponding ones in the
l.h.s. of~(\ref{N(z)}) we come to the relations
\begin{align}
&\varphi(1) \!+\! g(\beta)\rho(1) = 1,\\
&\varphi(2) \!+\! g(\beta)\rho(2) = 2 (1 - 2^{7/2}a/\lambda_T).
\end{align}
From these we find
\begin{align}
&g(\beta) = 2^{9/2}(a/\lambda_{T_0}-a/\lambda_{T}) \rho^{-1}(2),\\
&\varphi(1)=1,\ \ \rho(1)=0,\nonumber
\end{align}
where $T_0$ is defined from $\varphi(2) = 2 (1 -2^{7/2}a/\lambda_{T_0})$.

The first example of the respective deformed derivative is
\begin{equation}\label{Der_ex1}
\Bigl[z\frac{d}{dz}\Bigr]_q + (\lambda_{T_0}/\lambda_T-1) (q-1)\Bigl(z\frac{d}{dz}-1\Bigr)
\end{equation}
where
\begin{equation}\label{q_ex1}
q=1-2^{9/2}a/\lambda_{T_0}.
\end{equation}
Note, the form of the latter is very natural: it shows that
the extent (magnitude) $1-q$ of deformation is just proportional
to the scattering length $a$ divided by $\lambda_{T_0}$.

More general case is the $\tilde{\mu},q$-deformed one. For it,
\begin{align}
&z\tilde{\mathcal{D}}_z = \varphi_{\tilde{\mu},q}\Bigl(z\dd{}{z}\Bigr) + 
(\lambda_{T_0}/\lambda_T-1) (q-1)\Bigl(z\frac{d}{dz}-1\Bigr) - \nonumber\\
&- 2 \frac{\sum_n \bigl(e^{-2\beta_{T_0'} \varepsilon_n^{int}} - 
e^{-2\beta_T \varepsilon_n^{int}}\bigr)}{1 - \sum_n e^{-2\beta_{T_0'} \varepsilon_n^{int}}} 
\tilde{\mu} \Bigl(z\frac{d}{dz}-1\Bigr), \label{Der_ex2}\\
&q=1-2^{9/2}a/\lambda_{T_0},\quad \tilde{\mu}=1 - \sum\nolimits_n e^{-2\beta_{T_0'} \varepsilon_n^{int}}.\label{muq_ex2}
\end{align}
The comparison of eqs. (\ref{q_ex1}) and (\ref{muq_ex2}) shows the difference
between the two situations. In the former, only interaction
is effectively taken into account, while in the latter, more general, 
case the both factors -- interaction and compositeness -- are involved.

Remark that besides modified deformed  derivative~(\ref{modif}),
 its further extensions may be considered, e.g.
\begin{equation}\label{modif2}
z\tilde{\mathcal{D}}_z\equiv \varphi\Bigl(z\dd{}{z}\Bigr) + \chi\Bigl(z\dd{}{z}\Bigr) h\Bigl(\dd{}{\beta}\Bigr) + g(\beta) \rho\Bigl(z\dd{}{z}\Bigr) + ...\,.
\end{equation}
For the commutator $[h(\partial/\partial\beta),f(z\partial/\partial z)]$ we obtain
\begin{equation}
\Bigl[h\Bigl(\dd{}{\beta}\Bigr),f\Bigl(z\dd{}{z}\Bigr)\Bigr] = \sum_{k=1}^\infty \frac{\beta^{-k}}{k!} Q_k\Bigl(z\dd{}{z}\Bigr) h^{(k)}\Bigl(\dd{}{\beta}\Bigr)
\end{equation}
where $Q_k(x) \equiv (-1)^k x^{-k}\Bigl(x^2\frac{d}{dx}\Bigr)^k
f(x)$. So, since the ambiguity in the position of
$h\bigl(\dd{}{\beta}\bigr)$ is still present, the terms with higher
derivatives $h^{(i)}\bigl(\dd{}{\beta}\bigr)$ may enter the
``$...$'' in~(\ref{modif2}).

\section{Concluding remarks}

In the analysis of the second virial coefficient of non-ideal Bose
gas from the viewpoint of the role of such important factors of
non-ideality as interaction of particles and their compositeness, we
have found explicit expression for $V_2-V_2^{(0)}$ given through the
scattering length $a$ and the effective radius $r_0$ of interaction.
The latter result, when compared to virial coefficient (\ref{V(q)})
of deformed Bose gas, has led us to one of our main formulas, eq.~(\ref{q(a,r,T)}). 
Though the dependence of deformation parameter on
$a$ and $r_0$ is rather expected, the encountered $T$-dependence has
become a kind of surprise. With the goal to find reasonable
explanation, and to justify the appearance of $T$-dependence in
$q$, we developed the appropriate extension of the very starting
point of the procedure to ``deform'' thermodynamics. Main step consists
in adopting the modified derivative eq.~(\ref{modif}) or its more
general version~(\ref{modif2}) involving additional structure functions.

In a similar way, when we deal with compositeness and derive the
 relation (\ref{mu(E,T)}) for $\tilde{\mu}=\tilde{\mu}(\varepsilon_n^{int},
\Phi^{\mu\nu}_{\alpha}, T)$, the appearance of temperature
dependence in the effective description can be described by usage of
eq.~(\ref{modif}) as well, however with different structure
functions involved.

Besides the above considered extension of the deformation, the possibility remains to
obtain another consistent deformation of a Bose gas with temperature dependent deformed
virial coefficients. The corresponding analysis is in progress.

As a further direction of research let us point out the unified microscopic treatment of the second virial
coefficient when the both factors of the compositeness and interaction are present simultaneously.
Note that the corresponding dependence $V_2=V_2(a,r_0,\varepsilon_n^{int},..,T)$ may be different
from that obtained above, and thus lead to some other structure functions of deformation.

\begin{acknowledgments}
The research was partially supported by the Special Program of the Division of Physics and
Astronomy of the NAS of Ukraine (A.M.G., Yu.A.M.), and by the Grant (Yu.A.M.) for Young Scientists
of the NAS of Ukraine (No. 0113U004910).
\end{acknowledgments}

\appendix
\renewcommand\thesection{\Alph{section}}

\section{Examples of the interaction potential}\label{ap1}

Here we present a number of examples of the interaction potentials
and respective phaseshifts or scattering length/effective radius
through which the second virial coefficient in (\ref{VirCoef}) or (\ref{VirCoef2}) is expressed.

\paragraph{Hard spheres interaction potential.} \label{HS-par}
It is given by
\begin{equation}\label{HSpheres}
U(r) = \left\{
\begin{aligned}
+&\infty,\ &r<D;\\
&0,\ &r>D.
\end{aligned}
\right.
\end{equation}
Direct calculation using~(\ref{VirCoef}) yields (see e.g.~\cite{Pathria})
\begin{equation}
V_2-V_2^{(0)} = \left\{
\begin{aligned}
&2\frac{D}{\lambda_T} + \frac{10\pi^2}{3} \Bigl(\frac{D}{\lambda_T}\Bigr)^5 + ... \\
&\qquad(\text{Bose case},\ l=0,2);\\
&6\pi \Bigl(\frac{D}{\lambda_T}\Bigr)^3 - 18\pi^2 \Bigl(\frac{D}{\lambda_T}\Bigr)^5 +...\\
&\qquad(\text{Fermi case},\ l=1).
\end{aligned}
\right.
\end{equation}

\paragraph{Constant repulsive potential.}
It is defined by
\begin{equation}\label{CR-pot}
U(r) = \left\{
\begin{aligned}
&U_0>0,\ &r<R;\\
&0,\ &r>R.
\end{aligned}
\right.
\end{equation}
Respective $l=0$ phaseshift and scattering length are then given as (for this 
and further examples see e.g.~\cite{Flugge1999})
\begin{equation}
\delta_0 = kR \Bigl(\frac{{\rm th}\, K_0R}{K_0R}-1\Bigr),\ \ \ a = R \Bigl(1-\frac{{\rm th}\, K_0R}{K_0R}\Bigr)
\end{equation}
where $K_0^2 = \frac{2m U_0}{\hbar^2}$,  and $r_0=0$ (see also eq. (\ref{a,r_0-f})).
The difference $V_2-V_2^{(0)}$ can be calculated using (\ref{VirCoef2})
like in the previous case (this concerns also the rest of examples).

\paragraph{Square-well potential.} It has the definition:
\begin{equation}\label{SqW}
U(r) = \left\{
\begin{aligned}
-&U_0<0,\ &r<R;\\
&0,\ &r>R.
\end{aligned}
\right.
\end{equation}
 The scattering length and effective radius equal to
\begin{equation}
a = -R \Bigl(\frac{\tg K_0R}{K_0R}-1\Bigr),\ \ r_0 =
R \Bigl(1-\frac{1}{K_0^2Ra}-\frac{R^2}{3a^2}\Bigr)
\end{equation}
with $K_0$ defined as in the previous example.

\paragraph{Anomalous scattering potential.} In this case
\begin{equation}\label{AnomScat}
\frac{2m}{\hbar^2} U(r) = \left\{
\begin{aligned}
-&K_1^2,& &0\le r<r_1;\\
+&K_0^2, & &r_1\le r\le R;\\
&0, & &R\le r.
\end{aligned}
\right.
\end{equation}
For $l=0$ phaseshift we have
\begin{equation}
\delta_0 = -kR + \arctg \biggl\{\frac{kR}{\kappa R} \cdot \frac{{\rm th}\,
\kappa(R\!-\!r_1) + \kappa r_1 \frac{\tg Kr_1}{Kr_1}}{1
+ \kappa r_1 \frac{\tg Kr_1}{Kr_1} {\rm th}\, \kappa(R\!-\!r_1)}\biggr\}
\end{equation}
where $\kappa^2 = K_0^2-k^2$, $K^2 = K_1^2+k^2$.

\paragraph{Scattering resonances.} The corresponding potential is
\begin{equation}\label{SR_pot}
U(r) = \frac{\hbar^2}{2m} \frac{\Omega}{R} \delta(r-R).
\end{equation}
Phaseshift $\delta_0$ is given as
\begin{equation}
\tg (kR+\delta_0) = \frac{\tg kR}{1 + \Omega \frac{\tg kR}{kR}}.
\end{equation}

\paragraph{Modified P$\rm\ddot{o}$schl-Teller potential.}
The potential is
\begin{equation}\label{PT-pot}
U(r) = - \frac{\hbar^2\alpha^2}{2m} \frac{\lambda(\lambda-1)}{\ch^2 \alpha r}.
\end{equation}
The respective phaseshift $\delta_0$ reads:
\begin{multline}
\delta_0 = \arctg\frac{2\tilde{k}}{\lambda} - \arctg\bigl(\ctg\frac{\pi \lambda}{2} {\rm th}\, \pi \tilde{k}\bigr) +\\
+ \sum_{n=1}^\infty \Bigl\{\arctg\frac{2\tilde{k}}{\lambda+n} - \arctg\frac{2\tilde{k}}{n}\Bigr\},\ \ \tilde{k}=\frac{k}{2\alpha}.
\end{multline}

\paragraph{Inverse power repulsive potential\,,} that is
\begin{equation}\label{NegPow-pot}
U(r) = \frac{\hbar^2}{2m} \frac{g^2}{r_0^2} \Bigl(\frac{r_0}{r}\Bigr)^n.
\end{equation}
For it, the scattering length and the wavefunction of the lowest state (through which
the effective radius is expressed) are respectively given as
\begin{align}
&a = r_0 \frac{\Gamma(1-\frac{1}{2\eta})}{\Gamma(1+\frac{1}{2\eta})} \Bigl(\frac{g}{2\eta}\Bigr)^{\frac{1}{\eta}}, \
\eta= \frac{n-2}{2},\\
&\chi_0 = C \sqrt{\frac{r}{r_0}} K_{\frac{1}{2\eta}} \Bigl(\frac{g}{\eta} (r/r_0)^{-\eta}\Bigr),\label{NP-chi0}
\end{align}
$K_\nu(z)$ being the modified Hankel function.

\paragraph{First Born approximation.}
Finally, let us present the expression for the $l$th phaseshift in the first Born
approximation:
\begin{equation}
\delta_l \simeq -\frac{2mk}{\hbar^2} \int\limits_{0}^\infty U(r) \bigl(j_l(kr)\bigr)^2 r^2dr
\end{equation}
where $j_l$ is the spherical Bessel function. However its applicability is quite restricted,
and the respective validity conditions reduce to the ones on $U(r)$ of the first Born approximation.

\bibliographystyle{apsrev4-1}
\bibliography{references}

\end{document}